# Phase diagrams of $La_{2-x}Sr_xCuO_4$ and $YBa_2Cu_3O_{6+\delta}$ as the keys to understanding the nature of high-$T_c$ superconductors

**K.V. Mitsen, O.M. Ivanenko**

*Lebedev Physical Institute RAS, 119991 Moscow, Russia*

**Abstract**
1. **Introduction**
2. **The electronic structure of high-$T_c$ superconductors**
3. **Doping and the formation of active NUC's**
4. **The formation of NUC's in $La_{2-x}Sr_xCuO_4$**
5. **Ordering of dopant ions and percolation in $La_{2-x}Sr_xCuO_4$**
6. **Phase diagram of $La_{2-x}Sr_xCuO_4$.**
7. **Transformation of electron structure of $La_{2-x}Sr_xCuO_4$ under doping**
8. **Mechanism of hole carrier relaxation**
9. **The formation of NUC's in $YBa_2Cu_3O_{6+\delta}$**
10. **Fluctuations and the nature of pseudogap. Phase diagram of $YBa_2Cu_3O_{6+\delta}$**
11. **Stripes**
12. **Conclusion.**
13. **References**

**Abstract**

The model that explains many of the details of superconducting and magnetic phase diagrams of $YBa_2Cu_3O_{6+\delta}$ and $La_{2-x}Sr_xCuO_4$ is presented. The model is based on the assumption of rigid localization of doped charges in the close vicinity of doped ion. This localization results in local variation of electronic structure of the parent charge-transfer insulator that depends on the local mutual arrangement of the doped charges. It is shown that in such system the negative-U centers (NUCs) may form under certain conditions on the pairs of neighboring Cu cations in $CuO_2$ plane. We consider the mechanism of hole generation. The calculated dependences of hole concentration in $YBa_2Cu_3O_{6+\delta}$ on doping $\delta$ and temperature are found to be in a perfect quantitative agreement with experimental data. In the framework of the model the phase diagrams of $YBa_2Cu_3O_{6+\delta}$ is considered and the interpretation of pseudogap and 60 K-phases in $YBa_2Cu_3O_{6+\delta}$ is offered. The pseudogap has superconducting nature and arises at temperature $T^* > T_{c\infty} > T_c$ in small clusters uniting a number of NUCs due to large fluctuations of NUC occupation. ($T_{c\infty}$ and $T_c$ are the temperatures of superconducting transition for infinite and finite clusters, accordingly). The calculated $T^*(\delta)$ and $T_c(\delta)$ dependences are in accordance with experiment. The range between $T^*(\delta)$ and $T_c(\delta)$ corresponds to the range of fluctuations where small clusters fluctuate between superconducting and normal states owing to fluctuations of NUC occupation. The phase diagram of $La_{2-x}Sr_xCuO_4$ is calculated. It is shown the features of the superconducting phase diagram and the characteristics of stripe textures of $La_{2-x}Sr_xCuO_4$ only reflect the geometrical relations existing in a square lattice and the competition of different types of dopant ordering.

## 1. Introduction

In the 19 years that have passed since the discovery of high-$T_c$ superconductors [1] many different and mutually exclusive models have been suggested [2], models that to one extent or another explain the nature of the ground state and the anomalous properties of these compounds. However, the absence of a crucial experiment has made it impossible to choose the only correct model among the spectrum of models. Today, the most thoroughly studied high-$T_c$ superconducting compounds are $La_{2-x}Sr_xCuO_4$ and $YBa_2Cu_3O_{6+\delta}$, whose phase diagrams exhibit many well-reproducible features and have been thoroughly studied in the entire doping range. Hence, the results of comparisons between the experimental phase diagrams of $La_{2-x}Sr_xCuO_4$ and $YBa_2Cu_3O_{6+\delta}$ and the phase diagrams obtained in this or that model may serve as the key argument in selecting the correct mechanism responsible for the unusual properties of high-$T_c$ superconductors.

Here we present an elementary model of high-$T_c$ superconductor, which, nevertheless, provides a good picture of the main properties of such compounds and, in particular, describes all the characteristic features of the phase diagram of $La_{2-x}Sr_xCuO_4$ and $YBa_2Cu_3O_{6+\delta}$. We believe that such agreement between theoretical and experimental phase diagrams can be considered as proof of the validity of the proposed model of high-$T_c$ superconductors.

## 2. The electronic structure of high-$T_c$ superconductors

As is known, high-$T_c$ superconductivity is observed in cuprates and bismuthates. To be more definite, we will focus on cuprates, although everything said about cuprates is applicable to bismuthates ($Ba_{1-x}K_xBiO_3$ and $BaPb_{1-x}Bi_xO_3$). According to band theory, in undoped cuprate high-$T_c$ superconductors the upper antibonding band $\sigma^*_{x^2-y^2}$, formed by $Cu3d_{x^2-y^2}$ and $O2p_{x,y}$ states, is half-filled, which means that we are dealing with a metal, whose band structure and Fermi surface are those depicted in Fig. 1 (the hatched area represents filled states) [3,4]. However, it is a fact that these compounds in an undoped state are insulators, and the reason for this is the strong electron correlation on the Cu ion that obstructs the presence of two electrons on copper (the $d^{10}$ configuration). The electronic spectrum of the insulating phases of high-$T_c$ superconducting compounds near the Fermi level $E_F$ can be approximated by the model of a charge-transfer insulator [5], i.e., an insulator with a gap $\Delta_{ct}$ related to charge transfer. In this model (Fig. 2a), the empty upper Hubbard band, formed by unoccupied Cu3$d^{10}$ orbitals of copper ions in the CuO$_2$ plane, is separated from the filled lower Hubbard band by the repulsive energy of two electrons on copper, $U_H$. Inside the gap there is the filled band formed by oxygen $p_{x,y}$-orbitals. Hence, the gap $\Delta_{ct}$ in the spectrum is related to electron transfer from oxygen to the neighboring cation and amounts to about 1.5-2 eV for all high-$T_c$ superconductors. Within the simple ionic model, the size of $\Delta_{ct}$ is given by the following formula [6]:

$$\Delta_{ct} \sim \Delta_{EM} + A_p - I_d$$

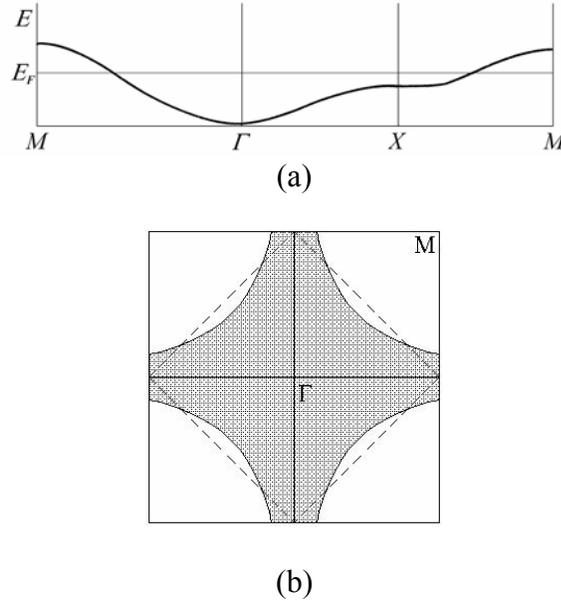

(b)

Fig. 1. (a) The electronic spectrum of the $CuO_2$ plane predicted by band theory, and (b) the corresponding Fermi surface (the hatched area corresponds to filled states). The dashed line indicates the boundary of the Fermi surface in the tight-binding model with allowance for nearest-neighbor interaction only.

Here $I_d$ is the second ionization potential of copper, $A_p$ is the electronegativity of oxygen in relation to $O^{2-}$ formation, and $\Delta_{EM}$ is the difference in the electrostatic Madelung energies between two configurations, in one of which the given copper and oxygen ions are in states $Cu^{2+}$ and $O^{2-}$ and in the other in states $Cu^+$ and $O^-$. Allowing for the fact that $I_d \sim 20$ eV and $\Delta_{ct} \sim 1.5$-2 eV, we see that the balance between these three quantities is very delicate. This situation can be changed or by heterovalent doping, e.g., by doping La$_2$CuO$_4$ with bivalent Sr or by doping Nd$_2$CuO$_4$ with tetravalent Ce, or by doping YBa$_2$Cu$_3$O$_6$ with excessive oxygen. What is important is that adding electrons (to Cu orbitals) or holes (to O orbitals) leads to the same result: a decrease in $\Delta_{EM}$ and, hence, in $\Delta_{ct}$. At a certain critical concentration the gap $\Delta_{ct}$ vanishes over the entire crystal and the substance becomes an ordinary metal. This is how the ionic model explains the transition of a charge-transfer insulator into a metallic state under doping. However, we believe that in high-$T_c$ superconductors the rigid localization of doped charges results in local variations of electronic structure of the parent charge-transfer insulator depending on the local mutual arrangement of doped charges.

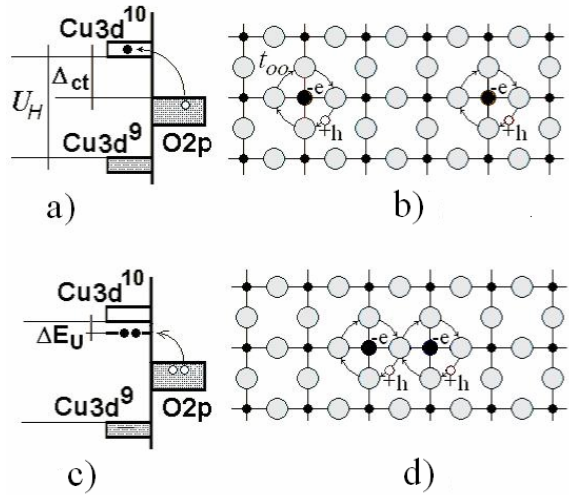

Fig. 2. (a) Electron spectrum of undoped cuprate high-$T_c$ superconductor. $U_H$ is the repulsive energy for two electrons on Cu ion. The charge transfer gap $\Delta_{ct}$ corresponds to the transition of electron from oxygen to nearest Cu ion with the origin of hole extended over four surrounding oxygen ions (Fig. b). (c) The energy of two such excitations can be lowered if two side-by-side ''hydrogen'' pseudo-atoms form ''hydrogen'' pseudo-molecule (Fig. 1d).

Earlier we proposed [7] the mechanism of NUC formation in high-$T_c$ superconductors. According to the model [7] the NUC are formed on pairs of neighboring Cu-ions in the CuO$_2$ plane. Figure 2a shows the electronic spectrum of undoped high-$T_c$ superconductor. Here the charge-transfer gap $\Delta_{ct}$ corresponds to the transition of electron from oxygen to nearest Cu ion. The originating hole is extended over 4 surrounding oxygen ions (Fig. 2b) due to overlapping of their $2p_{x,y}$ orbitals. This formation (3d$^{10}$-electron+2p-hole) resembles a hydrogen atom. We have shown that the energy of two such excitations can be lowered (Fig. 2c) if such two side-by-side pseudo-atoms form a pseudo-molecule (Fig. 2d). This is possible due to formation of a bound state (of the Heitler-London type) of two electrons and two holes that emerge in the immediate vicinity of this pair of Cu ions.

Thus, as a fact we deal with an intracrystalline hydrogen molecule, where the 3d$^{10}$ electrons on copper act as nuclei and holes on O2p$_{x,y}$-orbitals act as electrons, with the overlap between the orbitals ($t_{OO}$) creating the possibility of holes transferring from one oxygen atom to another. The analogy with an H$_2$ molecule is justified since the distance between cations in all high-$T_c$ superconductors (cuprates and bismuthates) amounts to about 3.7 ± 4.0 Å and is close to the product $R_0 \varepsilon_\infty$, where $R_0 \approx 0.8$ Å is the distance between the nuclei in an H$_2$ molecule, and $\varepsilon_\infty$ is the high-frequency dielectric constant, with $\varepsilon_\infty \sim 4.5$-5 for all high-$T_c$ superconductors [8]. That is, nature has actually `implanted' in high-$T_c$ superconductors the ability to form an `intracrystalline' hydrogen molecule. A lowering of the energy is possible, as it is in the H$_2$ molecule, only for the bonding orbital of a singlet hole pair. Additional lowering of the energy, $\Delta E_U$, caused by the transition of two electrons to neighboring cations, can be estimated in the case at hand as

$$\Delta E_U \sim \Delta E_{H_2} / \varepsilon_\infty^2 \approx 0.23 \text{ eV} ;$$

where $\Delta E_{H_2} \approx 4.75$ eV is the binding energy in an H$_2$ molecule. If through E(2), E(1), and E(0). we denote the energies of the ground state with, respectively, two electrons on a given pair of copper ions, one electron, and an empty pair, the above implies that E(2)+E(0)<2E(1). Thus, we can assume that a pair of neighboring copper cations in the CuO$_2$ plane is a NUC. In other words, in the energy spectrum of a charge-transfer insulator in the case of a high-$T_c$ superconductor there appears a pair level that is $\Delta E_U$ below the bottom of the upper Hubbard 3d$^{10}$ subband (Fig. 2c).

If we now decrease $\Delta_{ct}$ to the point where the gap disappears for two-particle transitions but remains for one-particle transitions, we arrive at a system in which some of the electrons belonging to the oxygen valence band effectively interact with pair states, or NUC's. We believe that the role of the doping is to activate possible NUC's.

## 3. Doping and the formation of active NUC's

It is usually assumed that in the same way as in ordinary metals, in doped high-$T_c$ superconductors the Coulomb potential of the dopant ion is screened at a distance of about 1 Å, in view of which the distribution of doped charges in the CuO$_2$ plane is homogeneous. This makes it possible to assume that on the whole the electronic structure of the crystal is homogeneous. At the same time, the experiment demonstrates a pattern that is just the opposite of the one above, namely, that the doped charges are localized on a scale of an order of the lattice constant. This conclusion follows from the results of measurements of X-ray absorption fine-structure (XAFS) spectra [9] and NMR spectra [10] in La$_{2-x}$Sr$_x$CuO$_4$. Such strong localization is probably caused by the weak screening of the impurity potential in the Mott insulator [11].

We base our reasoning on the assumption that at fairly low temperatures, the doped holes (electrons) are rigidly localized in the immediate vicinity of an impurity ion. More precisely, in $La_{2-x}Sr_xCuO_4$ a hole is localized in the $CuO_2$ plane on four oxygen ions belonging to an oxygen octahedron adjacent to Sr ion [9,10] (Fig. 3a). In $YBa_2Cu_3O_{6+\delta}$ doped holes are localized in $CuO_3$-chains, in oxygen sheets including excessive oxygen ions (Fig. 3f).

Let us consider the conditions under which a NUC is formed on the given pair of copper ions in the $CuO_2$ plane. For this purpose, we select a fragment of the crystal structure common to all cuprates with hole doping. This is the $Cu_2R_2O_n$ cluster, where the copper ions are "built" into the $CuO_2$ plane and R = Cu in the $CuO_2$ plane in $La_2CuO_4$, R = Cu in chains in $YBa_2Cu_3O_{6+\delta}$, and R = Bi in BSCCO. We assume that a NUC is formed in the $CuO_2$ plane on a pair of Cu ions if a hole formed as a result of doping (doped hole) is localized in each of the oxygen squares surrounding R ions (Fig. 3a,f). For $YBa_2Cu_3O_{6+\delta}$, this requirement means that a NUC on a given pair of Cu ions is formed when three consecutive oxygen sites are filled in the $CuO_3$ chain over these ions. For $La_{2-x}Sr_xCuO_4$, where localized doped holes lie in the $CuO_2$ plane (on four oxygen atoms of the oxygen octahedron adjacent to the strontium ion [9,10]), this requirement is fulfilled when the distance between the R ions (or between the projections of strontium ions onto the $CuO_2$ plane, which is the same) is $3a$ or $a\sqrt{5}$, where $a$ is the lattice constant in the $CuO_2$ plane (Fig. 3b,d).

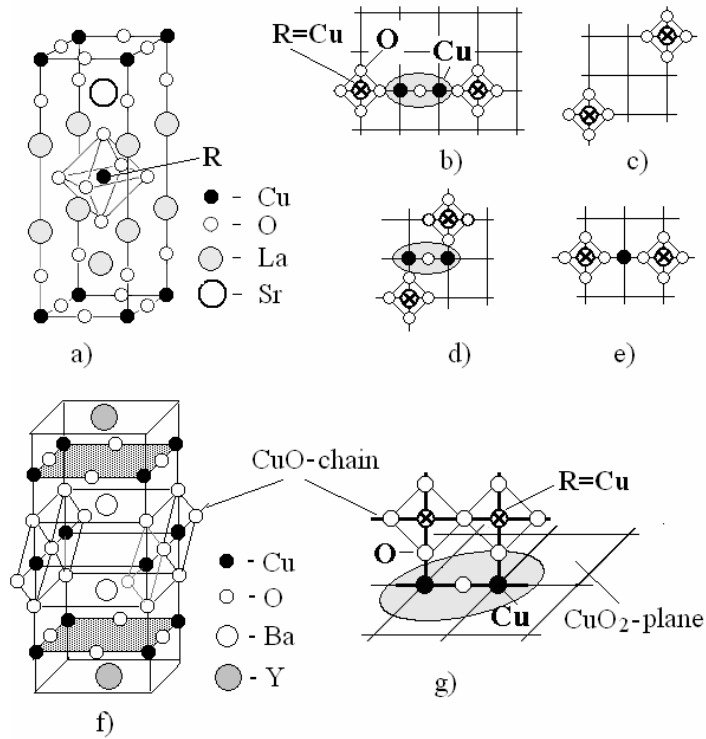

Fig.3. Formation of NUC in cuprates. (a) Unit cell of $La_{2-x}Sr_xCuO_4$. NUC (dashed oval) is formed on the interior pair of Cu ions in $CuO_2$ plane if the distance between R ions (see text) is $3a$ (Fig. b) or $a\sqrt{5}$ (Fig. d). In the intermediate case, when the distance between R ions is $a\sqrt{8}$ (Fig. c), there is no pair of neighboring copper ions with a doped hole adjacent to each of them, and NUCs are not formed. This situation corresponds to an insulator. When R ions are situated at a distance of $2a$ from each other (Fig. e), the $\Delta_{ct}$ gap for the internal copper ion is closed also for one-electron transitions. This corresponds to the conventional metal state.

## 4. The formation of NUC's in $La_{2-x}Sr_xCuO_4$

Let us check the validity of this statement for $La_{2-x}Sr_xCuO_4$. As was mentioned above, there can be two types of $Cu_2R_2O_n$ clusters satisfying our condition: with the distance between the R-ions (and doped holes) equal to either $3a$ or $a\sqrt{5}$ (Fig. 3b,d). The distance between the projections on the $CuO_2$ plane of the corresponding dopant ions is the same. The decrease in $\Delta_{ct}$ for Cu ions caused by the presence of a single hole near a given copper ion can be estimated by allowing only for the interaction between the nearest neighbors and assuming that this hole is `distributed' over the four nearest oxygen ions. The decrease in the energy of the $Cu3d^{10}$ state for four copper ions surrounding the doped hole is

$$\Delta E \approx e^2/4r \sim 1.8 \text{ eV}$$

(here, it is assumed that, being on an oxygen ion at a distance $r=a/2 \approx 2$ Å from the copper ion, the hole `sees' the unscreened copper ion; $e$ - is the electron charge). In other words, due to hole doping, the energy of these states is $\Delta E \approx 1.8$ eV below the bottom of the upper Hubbard band, which is roughly 0.1-0.2 eV smaller than the gap $\Delta_{ct}^0 \approx 1:9 \pm 2.0$ eV in undoped $La_2CuO_4$.

Thus, in both cases ($l=3a$ and $l= a\sqrt{5}$), the presence of a doped carrier in the vicinity of each R-ion decreases $\Delta_{ct}$ for the four neighboring copper ions and creates the conditions needed for tunneling in the inner copper ions of two $p_{x,y}$-electrons from the oxygen ions surrounding this pair and the forming of a bound state with the energy lowered by $\Delta_E \approx 0.23$ eV (without regards for hybridization). Here, the emerging singlet hole pair will be localized in the vicinity of the NUC. Thus, active NUC's are formed on fragments with $l=3a$ and $l= a\sqrt{5}$ (Fig. 3 b,d), and we believe such fragments to be the nuclei of a high-$T_c$ superconducting phase.

In the intermediate case, when the distance between the doped charges is $a\sqrt{8}$ (Fig. 3c), there is no pair of neighboring copper ions with a doped hole adjacent to each of them, and NUCs are not formed. This situation corresponds to an insulator and, as is shown below, is responsible for the "1/8"-anomaly. When doped holes are situated at a distance of $2a$ from each other (Fig. 3e), the $\Delta_{ct}$ gap for the internal copper ion is closed also for one-electron transitions. This corresponds to the conventional metal state.

Thus, doped carriers in our model are localized, and they are responsible for the formation of active NUC's. These NUC's act as pair acceptors and generate additional hole pairs, which are also localized in the vicinity of the NUC. Conduction occurs in such a system if these regions of hole localization form percolation clusters in $CuO_2$ plane and by means of quantum tunneling between such clusters. As well, conductivity can be provided if isolated NUC's are imbedded in metal matrix.

Pair hybridization of oxygen $p_{x,y}$-states with NUC states determines the behavior of high-$T_c$ superconductors. Electron pairing in such system, responsible for high-$T_c$ superconductivity, emerges because of strong renormalization of the effective electron-electron interaction when scattering with intermediate virtual bound states of NUC's is taken into account. Simanek [12] was the first to propose such a mechanism, which was then repeatedly discussed in the literature as applied to various systems, including high-$T_c$ superconductors [13-20].

## 5. Ordering of dopant ions and percolation in $La_{2-x}Sr_xCuO_4$

As noted above, the proposed mechanism of the interaction between electrons and pair states is effective when there is percolation over regions of localization of hole pairs. This, in turn, is possible only if the existing NUC's form percolation clusters. In $La_{2-x}Sr_xCuO_4$ such clusters can be formed along broken lines with links (connecting doped holes) whose length is $l=3a$ or $l=a\sqrt{5}$. In the general case, where the dopant ions are randomly distributed, one can hardly expect extended clusters to form. However, as we wish to show, in $La_{2-x}Sr_xCuO_4$ the doped holes occupy the sites of the certain square lattices with lattice constants depending on $x$, which are sub-lattices of $CuO_2$ lattice. This results in the formation of different percolative broken lines with links whose length $l_{com}$ is commensurable with $a$. According to our consideration the broken lines with links $l_{com}=3$ or $\sqrt{5}$ form percolative clusters of NUC and correspond to the high-$T_c$ phases, the cluster with links $l_{com}=\sqrt{8}$ is the insulator and the cluster with links $l_{com}=2$ is conventional metal. Lattices with $l_{com}>3$ correspond to the insulator.

In the considered scheme of the doping the system of Sr ion and hole located in oxygen sheet, represents an electric dipole that interacts with other dipoles through long range Coulomb potential. In such systems the orientation interaction between dipoles arises, that results in alignment of these dipoles by opposite poles to each other. Based upon crystal structure of $La_{2-x}Sr_xCuO_4$ it is possible to assume that replacement La on Sr will occur so that arising dipoles form vertical chains (reminding crankshafts) extended along $c$-axis (Fig. 4). Such arrangement simultaneously removes a question: what of La(Sr)O planes (Fig. 3a) dopes hole in the central $CuO_2$ plane.

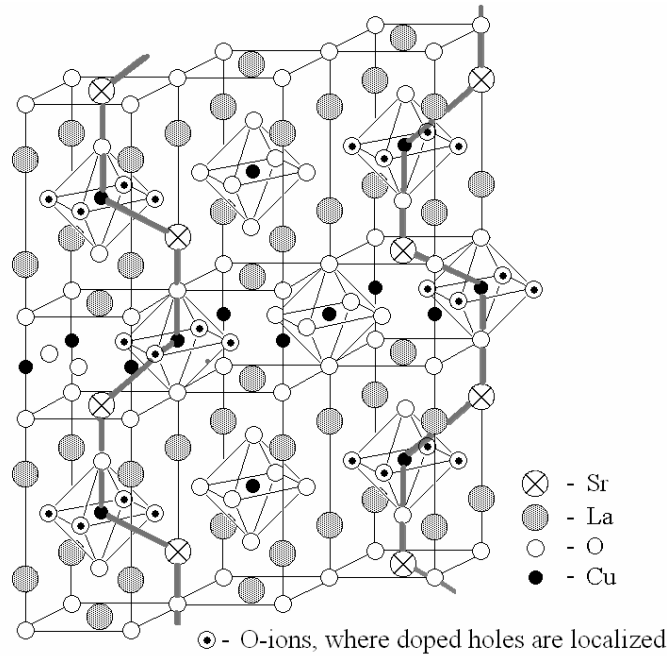

Fig. 4. Ordering of Sr ions in $La_{2-x}Sr_xCuO_4$. Negatively charged strontium ions together with doped holes "ascribed" to them are dipoles that attract each other at opposite ends to form cranked chains.

Let believe that chains are flat and drawn up parallel to each other. The calculation of electrostatic interaction energy of dipole chains shows, that two nearest chains (Fig. 4) will attract to each other if doped holes are spaced $l_{com}\geq 2$ and the next nearest chains repulse from each other. Such character of interaction results in dipole chain ordering and occupation of the sites in square lattice with certain parameter $l_{com}$ by doped holes. The minimum of interaction energy is appropriate to $l_{com}=\sqrt{8}$. At the same time it is follows from calculation that the energies of configurations with $l_{com}=2$, $\sqrt{5}$, $\sqrt{8}$ and 3 are close with a precision of $\sim 10^{-2} e^2/\varepsilon a$ per dipole ($\varepsilon$ - is the permittivity). Therefore the simultaneous coexistence of microdomains, in which doped holes occupy the sites in square lattices with definite but different $l_{com}$, is possible.

Microdomains with a given $l_{com}$ distance can only exist over a certain concentration $x$ range. This range is bounded from above by the $x_{com} = 1/l_{com}$ value; at higher concentrations, the existence of physically significant domains with given $l_{com}$ violates the condition of a constant mean concentration. At $x < x_{com}$, dipole chains become broken, and vacancies appear in the square lattices of projections. Microdomains with a given $l_{com}$ distance remain intact up to some $x = x_p$ value, which, at a random distribution, corresponds to the two-dimensional percolation threshold $x_p=0,593$[5]. That is the percolative cluster serves as a carcass of a given microdomain. Accordingly, the existence of microdomains with given $l_{com}$ is possible in the concentration range satisfying a condition

$$0,593/l^2_{com} < x \leq 1/l^2_{com}$$

The concentration of occupied sites $p$ in such microdomain changes with $x$ from $p\approx 0,6$ (at $x=0,593/l^2_{com}$) up to $p=1$ (at $x=1/l^2_{com}$).

The size of the ordered microdomain depends on the proximity of $x$ to $x_{com}$ and grows up to 200-600Å in **ab**-plane at x→0.12[6]. Along the **c**-axis the size of the ordered microdomain appears to be of some lattice constants, with the type of ordering of doped holes being repeated in every second $CuO_2$ plane.

We assume that, at small $x$ (at a mean distance between dopant projections of $l > 3a$), dipole chains are grouped in planes parallel to the **c** axis and the orthorhombic **a** axis in such a way that the distance between doped holes (or strontium ion projections) along the **a** axis be $a\sqrt{8}$, that is, correspond to minimum interaction energy.

## 6. Phase diagram of $La_{2-x}Sr_xCuO_4$.

As follows from our reasoning, $La_{2-x}Sr_xCuO_4$ must be treated as a set of mutually penetrating domains in which strontium ions are ordered in such a manner that doped holes fill (in part or completely) square lattice sites with various $l_{com}$ values determined by the concentration. The site percolation regions in lattices with various $l_{com}$ values, that is, the concentration regions corresponding to the existence of clusters of various phases, can be determined. Table 1[1] lists the existence intervals of domains with different $l_{com}$.

---

[1] Table 1. Intervals of the existence for microdomains with various $l_{com}$ values

| $l_{com}$ | $x_p$ | $x_m$ | properties |
|---|---|---|---|
| >3 | | | insulator |
| 3 | 0.066 | 0.111 | high-$T_c$ superconductor |
| $\sqrt{8}$ | 0.075 | 0.125 | insulator |
| $\sqrt{5}$ | 0.12 | 0.20 | high-$T_c$ superconductor |
| 2 | 0.15 | 0.25 | normal metal |

Note: $x_p$ and $x_m$ are the lower and upper boundaries of the concentration range in which domains with the given $l_{com}$ can exist, $x_m = 1/l^2_{com}$ ; $x_p = 0.593 x_m$ is the percolation threshold, when the existence of percolation chains with $l = l_{com}$ becomes possible. The last column contains characteristics of microdomains with the given $l_{com}$ value.

Figure 5a shows the intervals of concentrations corresponding to two-dimensional site percolation in domains with $l_{com} = 3$, $\sqrt{8}$, $\sqrt{5}$ and 2, that is, the intervals where, according to our reasoning, two-dimensional clusters with $l_{com} = 3$ and $\sqrt{5}$ (NUC chains), a cluster with $l_{com} = 2$ (conventional metal) and a cluster with $l_{com} = \sqrt{8}$ (insulator) can exist. The boundaries of the regions of the existence of percolation broken lines with segment lengths $l_{com}$ are shown by solid lines in Fig. 5a. The percolation regions for domains with $l_{com} = 3$ and $\sqrt{5}$, that is, for NUC chains, are indicated by thick lines. Figure 6 shows how the regions of the localization of singlet hole pairs overlap along percolation clusters with (a) $l_{com} = 3$ and (b) $l_{com} = \sqrt{5}$. Note that, as shown in Fig. 6a, in microdomains with $l_{com} = 3$ the motion of carriers occurs mostly along Cu–O bonds. This is in agreement with the results reported in [21], where the conclusion of such a character of the movement of carriers was drawn from the APRES, IR, and Raman data on $La_{1.9}Sr_{0.1}CuO_4$ crystals.

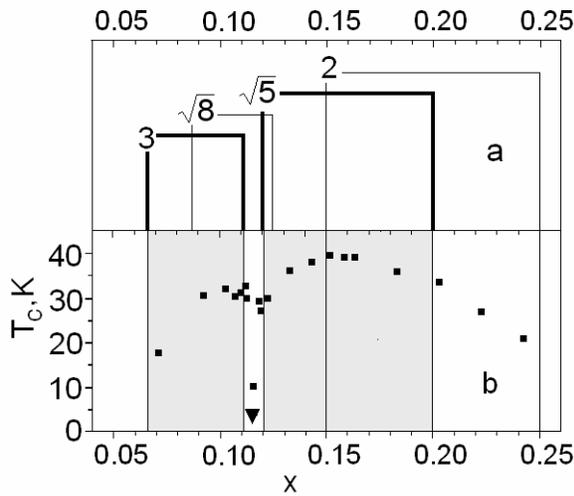

Fig. 5. (a) Intervals of concentrations corresponding to two-dimensional site percolation in domains with $l_{com} = 3$, $\sqrt{8}$, $\sqrt{5}$ and 2, that is, the intervals where two-dimensional clusters with $l_{com} = 3$ and $\sqrt{5}$ (NUC chains), a cluster with $l_{com} = 2$ (conventional metal) and a cluster with $l_{com} = \sqrt{8}$ (insulator) can exist. The boundaries of the regions of the existence of percolation broken lines with segment lengths $l_{com}$ are shown by solid lines. The percolation regions for domains with $l_{com} = 3$ and $\sqrt{5}$, that is, for NUC chains, are indicated by thick lines. (b) Solid squares are the experimental $T_c(x)$ phase diagram for $La_{2-x}Sr_xCuO_4$ from [22] The composition ($x$=0.115) for which superconductivity was not observed to 4.2 K is marked by a solid triangle.

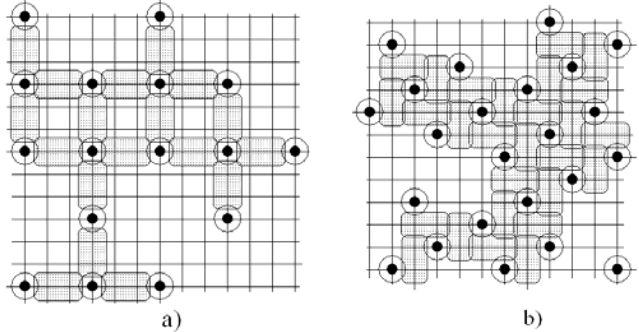

Fig. 6. Percolation clusters in the regions of the localization of singlet hole pairs: (a) $l_{com} = 3$ and (b) $l_{com} = \sqrt{5}$. Solid circles are the projections of dopant ions onto the $CuO_2$ plane. Open circles are doped hole localization regions, and rectangles are the localization regions of hole pairs around NUC's.

As illustrated in Fig. 5a, bulk superconductivity exists in the regions $0.066 < x < 0.11$ and $0.12 < x < 0.20$. In the first interval, corresponding to the undrdoped region, the bulk superconductivity arise due to Josephson coupling between superconducting microdomains with $l_{com} = 3$. In the region $0.12 < x < 0.15$ the only percolation over the clusters with $l_{com} = \sqrt{5}$ is possible. This range is corresponded to the optimal doping. At $0.15 < x < 0.20$ superconducting domains (containing clusters with $l_{com} = \sqrt{5}$) and normal metal domains coexist and the fraction of normal domains increases as $x$ grows. This corresponds to the transition to the overdoped state in which superconductivity is fully determined by the proximity effect and $T_c$ monotonically decreases as $x$ increases.

Figure 5(b) shows the experimental phase diagram $T_c(x)$ for $La_{2-x}Sr_xCuO_4$ from [22] The coincidence of the regions of superconductivity in the experimental phase diagram with intervals of percolation for $l_{com} = \sqrt{5}$ и 3 proves the conclusion that the fragments in question, which include pairs of neighboring Cu ions in $CuO_2$ plane together with two neighboring doped holes, are responsible for superconductivity in $La_{2-x}Sr_xCuO_4$. Note that the "dip" in $T_c(x)$ in the range $0.11 < x < 0.12$, caused by the absence of percolation over the chains of NUC's is superimposed on the region of existence (as $x \to 1/8$) of $\sqrt{8} \times \sqrt{8}$ lattice of doped holes corresponding to the insulating phase. Further we show that just this feature makes it possible to observe a static incommensurable magnetic texture in this region.

As follows from above consideration at $x > 0.066$ there can be two-electron transitions to some pairs of neighboring copper ions. The experiments corroborate this assertion very vividly. Figure 7 shows the dependence of the hole carrier concentration $n_{f.u.}$ (per formula unit) on $x$ for $La_{2-x}Sr_xCuO_4$ [23]. Clearly, beginning with $x > 0.05$, when, according to our model, clusters of NUC's with $l_{com} = 3$ emerge, the function $n_{f.u.}(x)$ deviates from unity, which suggests that new carriers appear in addition to doped carriers.

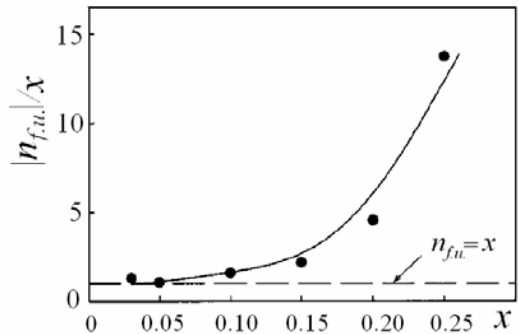

Fig. 7. Dependence of the hole carrier concentration $n_{f.u.}$ (per formula unit) on $x$ for $La_{2-x}Sr_xCuO_4$ (the data was taken from Ref. [23]

## 7. Transformation of electron structure of $La_{2-x}Sr_xCuO_4$ under doping

As was mentioned above, undoped high-$T_c$ superconductors with a half-filled band are insulators, although according to simple band theory, which ignores correlation effects, they should have been metals. Let us examine, in a qualitative manner, how the transition from a metallic state to an insulating state occurs when electron-electron correlations are taken into account.

Koster and Slater [24] examined the change in the band structure of a crystal caused by the implantation of an isolated defect into the crystal. Their main result was that in the one-dimensional case the implantation of a single defect characterized by an additional local potential $U$ with respect to the undistorted periodic potential leads to a situation in which a single state separates from the band in question. If $U < 0$, it is the lower state that splits off, while if $U > 0$, it is the upper state that splits off. The separated state is localized in the near vicinity of the defect. Only insignificant shifts of states within the band emerge as a result, and the wave function within the band remains a delocalized Bloch function.

Now let us `switch on' the electron-electron correlations on the copper ions one after another. We begin by increasing the potential on a single copper ion by $U$. Then one upper state, corresponding to a certain **k**, splits off from the half-filled band. If we `switch on' the correlations on all copper ions, the filled states (formed primarily by oxygen orbitals) will find themselves separated from the unoccupied states (mainly copper orbitals) by a gap. As a result, on a certain contour surrounding the point M$(\pi,\pi)$ (Fig. 1) there appears a gap $\Delta_{ct}$, i.e., the spectrum begins to resemble that of an insulator.

If we ignore the shift of states, the electronic spectrum near $E_F$ can be represented in the form of two bands: one filled O2p$_{x,y}$-band, whose shape in momentum space resembles that of a volcano (Fig. 8), and one unoccupied band (shaped like a hat), formed primarily from localized Cu3d$^{10}$-orbitals. When we take into account the shift of states, the O2p$_{x,y}$-band in the insulator phase will be deformed, but the projection of the band pattern on the ($k_x$, $k_y$) plane (precisely, the contour along which the occupation number $n(k)$ diminishes substantially) for an insulator will, to a certain extent, retain the shape of the contour of the initial Fermi surface of the conductor in the absence of correlations, which follows from the fact that the boundary of the region of filled states in the direction of the point M$(\pi,\pi)$ in the initial Fermi surface lies below the point $(\pi/2, \pi/2)$. This follows from the results of calculations of the energy bands for the CuO$_2$ plane within the three-band Hubbard model [25], which takes into account the overlap of p-orbitals of the nearest oxygen atoms. As a result, the maximum band width will be in the direction of **k** along the $(\pi,\pi)$ diagonal, i.e., along the chains of oxygen atoms, because of the overlap of oxygen 2p$_x$- and 2p$_y$-orbitals (the overlap integral $t_{pp}$), while the minimum band width of order $t_{pd}^2/U$ ($t_{pd}$ is the overlap integral of Cu 3d$_{x^2-y^2}$ and O2p$_{x,y}$-orbitals) will be in the direction $(0, \pi)$. Thus, the insulating gap will be of a d-wave form: with a minimum along the $(\pi,\pi)$ direction and a maximum along $(0,\pi)$ and $(\pi,0)$. The above picture agrees with the results of observations of the remnant Fermi surface [26].

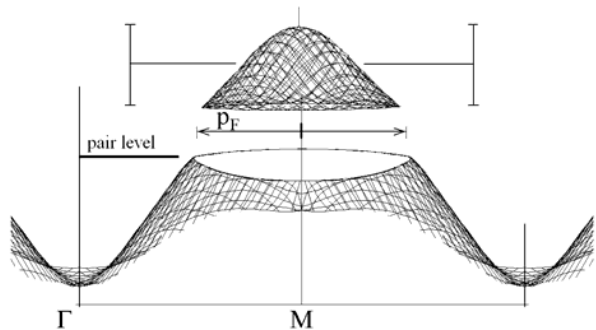

Fig. 8. Dielectrization of the spectrum of the CuO$_2$ plane (without allowance for the shift of states) due to correlations. The upper unoccupied states are localized in the vicinity of Cu ions. The lower filled states are formed primarily from oxygen p$_{x,y}$-orbitals. The insulating gap emerges on the contour of the Fermi surface that would have existed if there were no correlations.

Under doping of $La_{2-x}Sr_xCuO_4$ (when La is replaced by Sr), the additional doped holes are localized (at least at low temperatures) in the vicinity of an impurity ion and occupy states inside the gap. As a result, the region of filled electron states diminishes as $x$ grows, while the size of the region of hole states increases with $x$ as $(1+x)$.

The formation of a percolation cluster from NUC's, whose pair level descends below the top of the oxygen band, opens the possibility for two-electron transitions to occur between oxygen ions and the copper ions and for two-particle hybridization of this pair level with the band states that have the highest energy to take place. This is accompanied by the restoration of the Fermi surface (or a part of this surface) in the sense of the constant-energy curve $E=E_F$ in the $(k_x, k_y)$ plane, in which the occupation numbers $n(E)$ suddenly decrease (at $T=0$) from 1 to 0 as $E$ grows. However, in contrast to the Fermi surface of an ordinary metal, where for $T>0$ electrons emerge in states with an energy $E>E_F$, in the case at hand for $T>0$ the emerging electron pairs occupy states with an energy $E<E_F$ (see Fig. 9). Accordingly, holes appear in this process along the contour of the `restored' part of the Fermi contour.

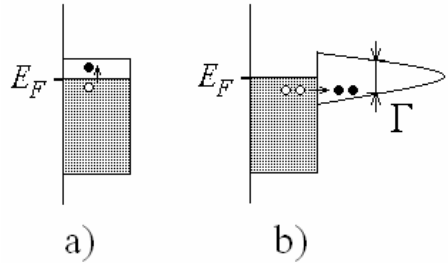

Fig. 9. (a) Band structure of a normal metal; (b) band structure of a high-$T_c$ superconductor. The holes appear in the filled band due to the transfer of electron pairs to a pair level whose width $\Gamma \propto T$.

When photoemission experiments are held for this case, a large Fermi surface of the hole type is observed. In the $La_{2-x}Sr_xCuO_4$ compound with $0.066<x<0.11$, due to the quasi one-dimensional nature of the network of percolation clusters from NUC's with $l=3a$ (Fig. 6a), the excitations in the $(\pi,\pi)$ direction are suppressed and the Fermi surface is essentially quasi one-dimensional. At $x>0.10$, when there is percolation along chains of NUC's with $l=a\sqrt{5}$ (Fig. 6b), the Fermi surface acquires a shape characteristic of the two-dimensional case (Fig. 1b).

As x increases, at $x>0.15$ there form percolation clusters from the `metallic' regions in which the distances between the projections of the dopants are smaller than $a\sqrt{5}$. For the states belonging to such a cluster the gap in the one-electron spectrum disappears. In this case, because tetravalent Sr replaces trivalent La, the band filling less than 1/2 and, correspondingly, the conduction is of the n-type.

Note that within the interval $0.066<x<0.11$ we have a state in which insulating and superconducting regions coexist, while at $x>0.15$ there is a spatially inhomogeneous state in which normal-metal and superconducting regions coexist. The described successive transformation of the electronic structure of $La_{2-x}Sr_xCuO_4$ with increasing x agrees well with the results of the photoemission studies of Ino et al. [27].

Another interesting aspect worth studying is the change in the behavior of the temperature dependence of the rate of relaxation of nuclear spin caused by doping. The results of numerous experiments show that the Korringa law $1/T_1T$=const (where $T_1$ is the relaxation time for a nuclear spin in the normal state) holds in cases of underdoping and optimal doping for $^{17}$O and does not hold for $^{63}$Cu [28]. At the same time, in overdoped La$_{2-x}$Sr$_x$CuO$_4$ samples this law holds for $^{63}$Cu, too [29]. This means that in overdoped samples, in contrast to underdoped and optimally doped samples, copper states contribute substantially to the density of electron states on the Fermi surface, which agrees with the proposed pattern of evolution of the electron states of high-$T_c$ superconducting compounds subjected to doping.

## 8. Mechanism of hole carrier relaxation

The transfer of electrons from oxygen ions to NUC's leads to the formation of additional hole carriers (Fig. 9). As a consequence of hybridization of the pair level of a NUC with band states, both band and localized pair states prove to be broadened. Taking into account two-particle hybridization, the broadening $\Gamma$ of the pair level [17,18] can be expressed as follows:

$$\Gamma \approx 4\pi kT \cdot (V/E_F)^2$$

(here $V \sim 1$ eV is the hybridization constant, $E_F \sim 0.5$ eV is the Fermi energy, and $T$ is the temperature). Hence, the broadening of the pair level $\Gamma \sim 10\text{-}50\ kT$.

The chemical potential of the hole pairs (measured from the position of the pair level) is always negative: $\mu_{pp} \approx -T/N$, where $N$ is the number of hole pairs on the pair level. Since $N \gg 1$, we can assume that $\mu_{pp} \approx 0$ and coincides with the position of the pair level. Between the band and the pair level dynamic equilibrium sets in, i.e., $\mu_{pp} = \mu_p \approx 0$, where $\mu_p$ is the chemical potential of holes in the `oxygen' band. Hence, when the pair level is at the top of the O2p-band, the distribution of holes in the band is nondegenerate at all temperatures.

The occupancy $\eta$ of the pair states, as well as the concentration $n$ of additional holes in the oxygen band, is determined by the equality of the rates of `band - pair level' transition and back. If $P$ is the concentration of NUC's, then $P\eta = n$. The rate of `pair level - band' transitions is $P\eta\Gamma \propto T\eta$. The rate of the reverse process is determined by the electron - electron scattering frequency and is proportional to $\Gamma^2(2-\eta) \propto T^2(2-\eta)$. Thus, we arrive at the expression

$$\eta = 2T/(T+T_0) \tag{8.1}$$

and $n = 2PT/(T+T_0)$, where $T_0$ is temperature–independent constant, that can be detected from Hall measurements.

Thus, due to the interaction between electrons and NUC's, the hole carrier distribution proves to be nondegenerate in the sense that the hole chemical potential $\mu_p \leq 0$ at all temperatures, while degeneracy requires that $\mu_p$ be positive.

Allowing for the nondegeneracy of the distribution (the absence of Pauli blocking) and the high hole concentration ($10^{21} - 10^{22}$ cm$^{-3}$), we can expect that the leading contribution to relaxation processes is provided by electron-electron scattering (in our case, the scattering of hole carriers on each other). However, since the interaction of two holes in a system with NUC's corresponds to effective attraction, this is not the ordinary Coulomb scattering. In the case at hand, the main electron-electron scattering mechanism is similar to the one [30] that operates in metals and alloys with strong electron-phonon interaction. In such substances, for electrons that are inside a layer of thickness $k\theta_D$ ($\theta_D$ is the Debye temperature) on the Fermi surface, the effective electron-electron interaction corresponding to attraction and related to the exchange of virtual phonons is much stronger than the screened Coulomb repulsion. Hence, the principal channel of electron-electron scattering in this case is also caused by the exchange of virtual phonons. The contribution of these processes [30] becomes significant when $T < \theta_D$. Here, the electron-electron scattering amplitude does not depend on the energy $E$ of the scattered particles at $E \ll k\theta_D$ and rapidly decreases at $E \sim k\theta_D$. When $E > k\theta_D$, only the Coulomb scattering contributes to the scattering amplitude. In experiments the contribution of the electron-electron scattering to the electrical resistivity $\rho$ ($\rho = AT^2$), exceeding the electron-phonon contribution, was observed in aluminum [31] at $T < 4$ K and in superconducting compounds with A15 structure [32] at $T < 50$ K. Here, the amplitude $A$ exceed the value calculated on the assumption that the scattering mechanism is of the Coulomb type by a factor greater than ten.

Thus, the main contribution to the relaxation of hole carriers in high-$T_c$ superconductors is provided by electron-electron scattering accompanied by the formation of an intermediate bound state on NUC's, which can be described as the exchange of a virtual boson with an energy $\Omega$. Since $\Omega \sim 0.2$ eV, the temperature interval where the contribution of such processes is essential extends to $T \sim 10^3$ K. The temperature dependence $\rho(T)$ in such a model can be derived from the Drude formula:

$$\rho = m^* \nu / n \cdot e^2$$

where $m^*$ is the effective hole mass and $\nu$ is the hole carrier scattering frequency. When $\Omega \gg E$, the scattering amplitude is independent of the particle energy $E$. Hence, the scattering frequency $\nu$ is determined by the hole concentration and a statistical factor in the scattering cross section, i.e., the volume of the phase space available for the scattering particles, which is proportional to $E_1 + E_2$ (here $E_1$ and $E_2$ are the energies of the scattered particles measured from the band top). Hence,

$$\nu \propto n (E_1 + E_2)$$

For DC conduction, $E_1 \sim E_2 \sim \Gamma \propto T$ and $\nu \propto nT$, with the result that $\rho(T) \propto T$. Such a dependence has been observed in experiments with optimally doped samples of $YBa_2Cu_3O_7$, $La_{2-x}Sr_xCuO_4$, $Bi_2Sr_2CaCu_2O_y$, etc.

The predominant contribution of the electron-electron interaction to the scattering processes also has an effect on the frequency and temperature dependences of the optical conductivity $\sigma_{opt}$:

$$\sigma_{opt} = (e^2 n/m^*) \cdot [\nu/(\omega^2 + \nu^2)]$$

where $\omega$ is the light frequency and $\nu$ is the `optical' relaxation frequency. For electron-electron scattering (at a concentration $n \sim 10^{22}$ cm$^{-3}$), the collision frequency $\nu \geq 10^{15}$ s$^{-1}$. Hence, for the IR range, $\nu \gg \omega$, and the formula for the optical conductivity becomes even simpler:

$$\sigma_{opt} = e^2 n / m^* \nu$$

For `optical' relaxation, $E_1 \sim \omega$, $E_2 \sim \Gamma \propto T$, and $\nu \propto n\omega$, when $\omega \gg \Gamma$ and $\nu \propto nT$ when $\omega \ll \Gamma$, which suggests that $\sigma_{opt} \propto \omega^{-1}$ (for $\omega \gg \Gamma$) and $\sigma_{opt} \propto T^{-1}$ (for $\omega \ll \Gamma$). These results agree fully with the data of different experiment [33,34].

## 9. The formation of NUC's in $YBa_2Cu_3O_{6+\delta}$

As follows from Chapter 3 the NUC is formed in $YBa_2Cu_3O_{6+\delta}$ on a given pair of Cu ions in $CuO_2$- plane when three in a row oxygen sites in CuO-chain over (under) this Cu pair are occupied (Fig.3 f,g). The concentration of such triplets at random distribution of oxygen ions in chains is equal to $\delta^3$ per unit cell.

Isolated triplet of oxygen ions in chain forms two NUC's, one NUC in each of two $CuO_2$ planes (Fig.10a). If the number of consecutive oxygen ions in a chain $N_O > 3$, only every second triplet can form separated NUC (without common Cu ions) in each of $CuO_2$ planes (Fig.10 b). So, we may take that for $N_O > 3$ each triplet forms one NUC but only in one $CuO_2$ plane (Fig.10 b).

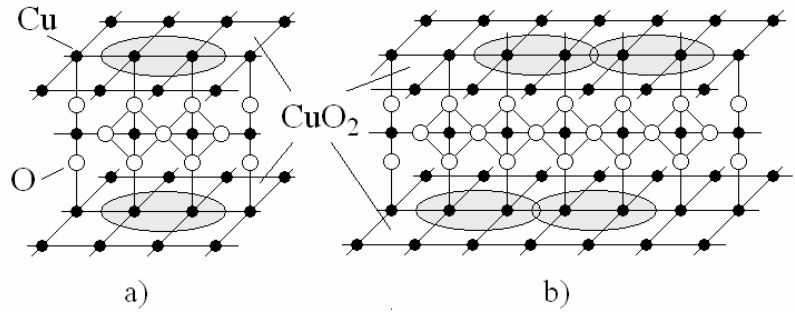

Fig.10. a) - the negative-U center (NUC) is formed in $YBa_2Cu_3O_{6+\delta}$ on a given pair of Cu ions in $CuO_2$ plane when three in a row oxygen sites in CuO-chain over (under) this Cu pair are occupied; b) formation of continuous NUC clusters in $CuO_2$ planes by the row of oxygen ions in CuO-chain.

We will consider that a few of NUC's lying along a straight line in $CuO_2$ plane belong to 1D-cluster if all Cu ions included in NUC's form continuous cluster of sites. Accordingly, oxygen ions in chains forming a given 1D-cluster of NUC's form a continuous oxygen 1D-cluster in a plane of chains. So, for each 1D-cluster of NUC's in $CuO_2$ plane there is continuous generative cluster of oxygen ions in CuO-chain. The continuous sets of oxygen ions belonging to adjacent chains we assume to form united 2D-cluster of NUC's provided that they touch over 3 or more oxygen ions in adjacent chains (that is the percolation over NUC's takes place). This will correspond to the formation of continuous 2D-clusters of NUC's in both $CuO_2$ planes. Figure 11 shows the random distribution pattern for oxygen ions in chains for the 40×40 square lattice obtained by this method for $\delta$=0.3 and $\delta$=0.6.

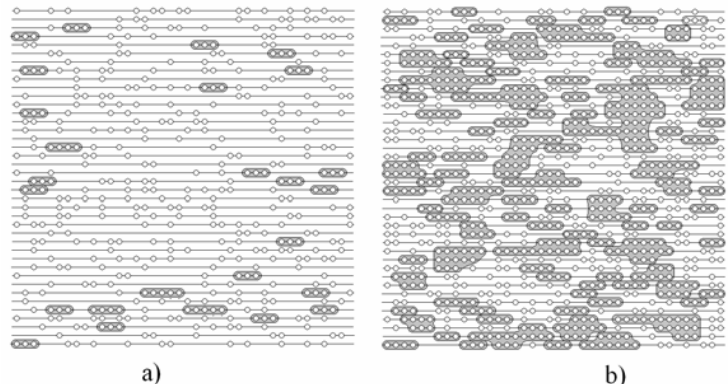

Fig.11. Pattern of chain oxygen clusters forming a finite clusters of NUC's in $CuO_2$ planes obtained by Monte-Carlo method for random distribution of oxygen ions in chains: (a) - $\delta$=0.3 and (b) - $\delta$=0.6. Open circles are oxygen ions in chains. Clusters contained more than 3 oxygen ions are dashed.

The total number of NUC's in clusters (for both $CuO_2$ planes) per one unit cell of $YBa_2Cu_3O_{6+\delta}$ at random distribution of oxygen ions is equal to $N_U = \delta^3 + N_3(\delta)$, where $N_3(\delta)$ is the $\delta$-dependent number of isolated triplets of oxygen ions in chains equal to $N_3(\delta)=\delta^3(1-\delta)^2$. Respectively,

$$N_U(\delta)=\delta^3\{1+(1-\delta)^2\} \qquad (9.1)$$

It is seen that $N_U(\delta) \approx \delta^3$ for $\delta > 0.65$, when the bulk of NUC's belongs to large clusters.

At $\delta < \delta_c$ NUC's form finite clusters of various sizes. Within each cluster the NUC occupation number $\eta$ depends on temperature and equals to $\eta = 2T/(T+T_0)$ (see Ch. 8).

As follows from (9.1) at $\delta > 0.65$ the volume concentration of NUC's $P = N_U/V_{UC} \approx \delta^3/V_{UC}$, where $V_{UC}=173 Å^3$ is the volume of unit cell for $YBa_2Cu_3O_{6+\delta}$. Accordingly, the volume concentration of hole carriers $n$, generated in $CuO_2$ planes as electrons occupied NUC is equal to $n=\eta P=\eta \delta^3/V_{UC}=2(\delta^3/V_{UC}) T/(T+T_0)$, and Hall coefficient is

$$R_H(\delta,T)=1\backslash ne=(1/2e)(V_{UC}/\delta^3)(T+T_0)/T, \qquad (9.2)$$

where $e$ is the electron charge. Figure 12a shows the temperature dependence of Hall coefficient for $YBa_2Cu_3O_{6.95}$ single crystals from [35], where authors first succeeded in separating of contributions from $CuO_2$ planes to Hall coefficient through the use of untwined $YBa_2Cu_3O_{6+\delta}$ single crystals with different $\delta$. As is seen from Fig. 12a the present data can be approximated successfully by (9.2) with $T_0 \approx 390$ K.

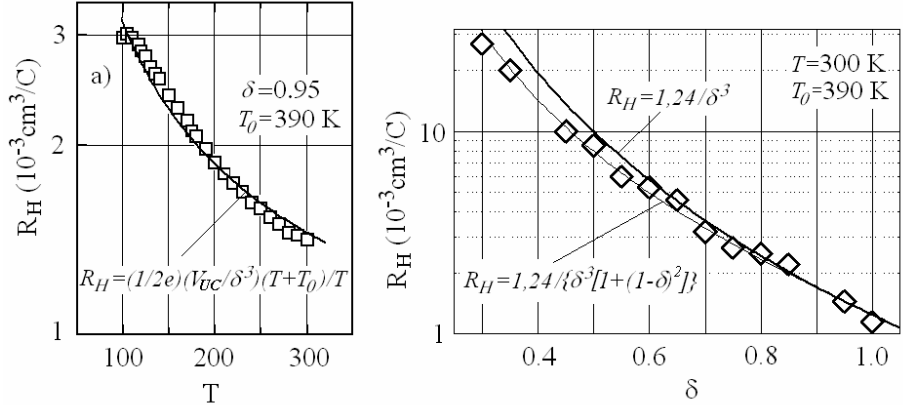

Fig.12 Hall coefficient of $CuO_2$ plane of twin-free $YBa_2Cu_3O_{6+\delta}$ single crystal depending on temperature and doping: (a) open squares - $R_H(T)$ for $\delta=0.95$ [35]; (b) open rhombuses - $R_H(\delta)$ at $T=300$K [35]. Solid line on both figures is the dependence (9.2) with $T_0=390$ K. Thin line is the $R_H(\delta)$ dependence with regard to the additional contribution of isolated triplets of oxygen ions in chains.

Fig. 12b shows experimental $R_H(\delta)$ dependence for $T=300$ K from [35]. It is seen that experimental data are in a good agreement with the dependence obtained from (9.2) with $T_0 \approx 390$ K for $0.65 < \delta < 1$. In order to get the $R_H(\delta)$ curve for the whole range of $\delta$ the relation (9.1) for $N_U$ should be used. The experiment (Fig. 12 b) completely confirms this conclusion. It should be emphasized that the calculated curves on the Figures 12 a,b have no any fitting parameter except $T_0$, that just describes the temperature dependence $R_H(T)$ and allows to calculate $R_H$ values over the whole range of $\delta$ and $T$ variation. That fact that hole concentration grows with a doping $\delta$ as $\delta^3$ can serve as powerful argument in favour of diatomic NUC existence in $YBa_2Cu_3O_{6+\delta}$ and as a support of the proposed mechanism of hole carrier generation in high-$T_c$ superconductors.

## 10. Fluctuations and the nature of pseudogap. Phase diagram of $YBa_2Cu_3O_{6+\delta}$

In [7] we have assumed that the pseudogap observed in different experiments is nothing but the same superconducting gap developing at $T>T_c$ due to the large fluctuations of NUC occupation because of electron transitions between the pair level of NUC and oxygen band.

The matter is in following. In conventional BCS superconductor with electron-phonon interaction the superconducting gap vanishes due to the thermal excitations over the Fermi surface, which decrease the number of unoccupied states available for the electron pair scattering. Analogously the mechanism of superconducting gap suppression in our case is the occupation of NUC's by real electrons. Therefore fluctuation-induced reduction of pair level occupation will amplify the superconducting interaction and can result in fluctuation-induced turning on superconductivity at $T^*>T>T_{c\infty}$ (here $T_{c\infty}$ is equilibrium value of $T_c$ for infinite cluster of NUC's). Opposite, the increasing of pair level occupation owing to fluctuation will reduce the superconducting interaction and can result in fluctuation-induced turning off superconductivity at $T_c<T<T_{c\infty}$. Large relative fluctuations of NUC occupation, corresponding to substantial deviation of $T^*$ and $T_c$ from $T_{c\infty}$ can happened in the underdoped samples when the significant number of NUC's belongs to finite nonpercolative clusters. The mean size of finite clusters decreases with doping reduction and relative fluctuations of NUC occupation increase in these clusters (i.e. $T^*$ goes up and $T_c$ goes down). In addition in overdoped samples when practically all Cu-ions belong to infinite percolation cluster, large fluctuations become impossible.

On the ground of the proposed model it is possible to deduce the dependences of $T^*$ and $T_c$ on doping $\delta$ for $YBa_2Cu_3O_{6+\delta}$. At $\delta<\delta_c$, when NUC's form finite clusters of various sizes, the sample should be defined as Josephson media, where superconductivity is realized over the whole volume thanks to the Josephson links between superconducting clusters.

The number of Cu ions included in cluster of NUC in $CuO_2$ plane we will take as a size of cluster, $S$. By the same time $S=3$ should be taken as a minimal cluster size where superconductivity could occur since it is impossible to consider the cluster with $S=2$ as superconducting due to the absence of superconducting transfer along cluster.

Let a cluster uniting some NUC's in $CuO_2$ plane contains $S \geq 3$ Cu-ions. According to (8.1), the number of electrons on NUC's in a given cluster at temperature $T$ is equal $N=TS/(T+T_0)$. This number can be changed on $\pm\sqrt{N} = \pm(TS/(T+T_0))^{1/2}$ due to fluctuations. The condition for turning superconductivity on (off) in a given cluster at temperature $T^*$ $(T_c)$ is $N(T) \pm \sqrt{N(T)} = N_c$, where $N_c=T_{c\infty}S/(T_{c\infty}+T_0)$ – the number of electrons on NUC at the superconducting transition temperature $T_{c\infty}$ for the infinite cluster. Hence,

$$TS/(T+T_0) \pm (TS/(T+T_0))^{1/2} = T_{c\infty}S/(T_{c\infty}+T_0). \qquad (9.3)$$

Solving Eq.(9.3) with $T_0$=390 K and $T_{c\infty}$=92 K we find $T^*$ and $T_c$ as a function of $S$ (Fig. 13). As it seen from Fig. 13, the fluctuation effect on $T_c$ decreases with cluster size increasing and becomes to be negligible in clusters of NUC with more than 1500 Cu ions, that corresponds to the size ~150Å. The "plateau" at 60K on the curve $T_c(\delta)$, where $T_c$ changes between 50 and 70K in the range $0{,}6<\delta<0{,}8$, corresponds to S changing from 10 to 100. It is notable that there is minimal $S$ value, below that the cluster does not remain superconductive even at $T\to 0$ due to fluctuations of NUC occupation. Since the NUC occupation is $\eta \approx 2/5$ at $T = T_{c\infty}$, any fluctuation in cluster with $S<5$ that increases the number of electrons on NUC by 2 will result in the destruction of superconducting state. In order to find $T^*(\delta)$ and $T_c(\delta)$ dependences it is necessary to know $\delta_c$ corresponding to the percolation threshold over NUC and the statistics of finite clusters of NUC depending on $\delta$.

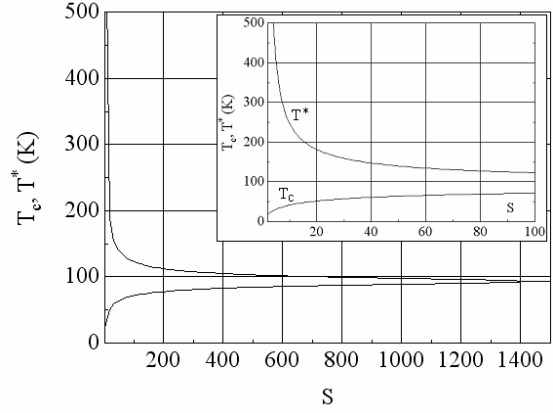

Fig.13. $T^*$ and $T_c$ as a function of cluster size $S$ for $3<S<1500$. Insert shows the same dependences for $3<S<100$. The "60-K plateau" on $T_c(\delta)$ curve where $T_c$ changes from 50 K to 70 K corresponds to $S$ variation on order of $S$ magnitude (from 10 to 100).

The percolation threshold over NUC as well as statistics of finite clusters can be found for the random distribution of oxygen atoms in chains by Monte Carlo method. According to the proposed mechanism of NUC formation we consider: 1) each 1D-cluster of oxygen ions in chain, containing $N_O \geq 4$ oxygen ions forms 1-D clusters of NUC with mean size $S = N_O - 1$ (i.e. containing $N_O - 1$ Cu-ions) in each of $CuO_2$ planes, 2) a size of 2D-cluster of NUC in $CuO_2$ plane is equal to the sum of sizes of constituent 1D-clusters. The value $\delta_c = 0,80 \pm 0,02$ has been determined by this method. It means that at $\delta > \delta_c$ $T_c$ would be equal $T_{c\infty}$. In the experiment however $T_c$ flattens out at $\delta > 0.85$ [36,37]. The increase of percolation threshold we suppose to connect with Cu-vacancies in chains and repulsion of oxygen atoms in adjacent chains [38], what prevents the growth of 1D-clusters. These factors will result in percolation threshold increasing in comparison with value expected for the random distribution of oxygen over the sites in chains.

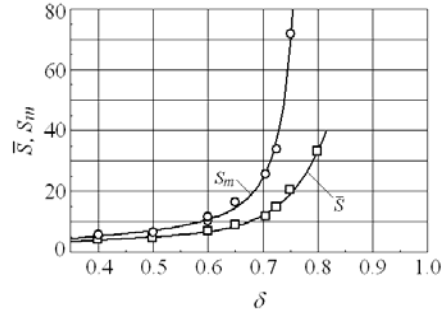

Fig.14. The mean sizes of finite clusters of NUCs, $S_m$ and $\overline{S}$ versus $\delta$ for $YBa_2Cu_3O_{6+\delta}$. Open circles and squares are the results obtained by Monte Carlo method for 40×40 lattice for $S_m$ and $\overline{S}$ correspondingly. Solid lines are drawn by eye.

To simplify the determination of $T^*(\delta)$ and $T_c(\delta)$ dependences we suppose that all finite clusters have the same sizes equal to some mean cluster size. The concept of mean cluster size $S_m$ is in use in the theory of percolation and is defined as weighted average $S_m = \sum n_i S_i^2 / \sum n_i S_i$. As it follows from the definition, large clusters give the basic contribution to $S_m$. And just this value $S_m(\delta)$ has to be substituted in (3) to find $T_c(\delta)$ dependence because $T_c$ is determined as a superconducting transition temperature of large clusters with higher $T_c$, shunting small clusters and governing the basic contribution to conductivity and diamagnetic response.

Opposite, in order to get $T^*(\delta)$ the simple average $\bar{S} = \sum n_i S_i / \sum n_i$ has to be used since the contribution to fluctuation-induced turning on superconductivity is determined by the all finite clusters proportionally to their sizes. Figure 14 shows $S_m(\delta)$ and $\bar{S}(\delta)$ for 40x40 lattice obtained by Monte Carlo method. Evidently that $S_m$ tends to infinity as the percolation threshold approaches while $\bar{S}$ remains finite at $\delta \geq \delta_c$. Substituting the obtained values of $S_m(\delta)$ and $\bar{S}(\delta)$ in the quadric equation (3) we get $T_c(\delta)$ and $T^*(\delta)$ for $YBa_2Cu_3O_{6+\delta}$ as two solutions of this equation. Both solutions are shown in Fig. 15 by triangles up and down, correspondingly. Solid lines are drawn by eye. **As follows from the model the area between these curves is the area of fluctuations, where the finite nonpercolative clusters fluctuates between superconducting and normal states due to fluctuations of occupations of NUC's.** The dotted line of $T_c(\delta)$ at $\delta<0.5$ corresponds to the area where the mean size of cluster of NUC's $\bar{S} <5$. As noted above, fluctuations will effectively destroy the superconductivity in these clusters. The experimental dependences $T^*(\delta)$ and $T_c(\delta)$ for $YBa_2Cu_3O_{6+\delta}$ single crystals are shown for comparison. Open squares are the data from [39], where the pseudogap opening temperature $T^*$ has been determined as the temperature of downward deviation of in-plane resistivity $\rho_{ab}(T)$ from the high-temperature $T$-linear dependence. Open rhombuses corresponds to the temperature of superconducting transition $T_c$, determined by magnetic measurements in [37]. As is seen from the comparison of experimental dependences $T^*(\delta)$ and $T_c(\delta)$ with the calculated ones the agreement can be considered as good in spite of convention of their definition.

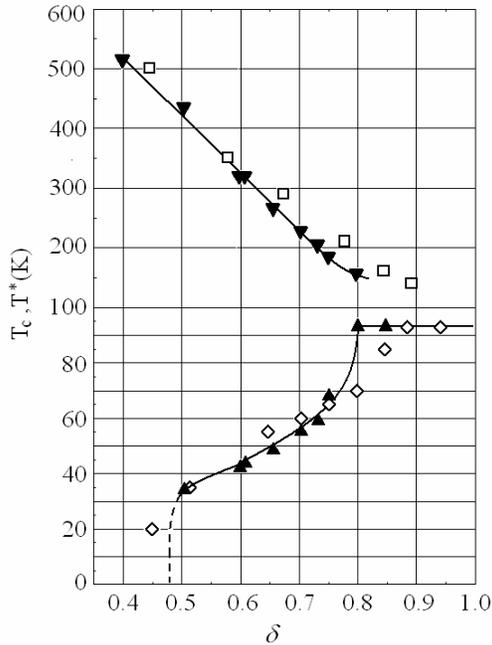

Fig.15. The comparison of calculated dependences of $T^*(\delta)$ (solid triangles down) and $T_c(\delta)$ (solid triangles up) for $YBa_2Cu_3O_{6+\delta}$. Open squares show the experimental results [39] for the single crystals, where $T^*$ has been determined as the temperature of downward deviation of in-plane resistivity $\rho_{ab}(T)$ from the high-temperature $T$-linear dependence. Open rhombuses are the results of the magnetic measurements of $T_c$ for $YBa_2Cu_3O_{6+\delta}$ single crystals [37]. Solid lines are drawn by eye. The dotted line of $T_c(\delta)$ at $\delta<0.5$ corresponds to the area where the mean size of cluster of NUC's $\bar{S} <5$ and fluctuations effectively destroy the superconductivity in these clusters.

## 11. Stripes

The incommensurable modulation [40-43] of AFM spin structure is observed in neutron scattering experiments [44] as two incommensurate peaks shifted in relation to AFM vector by $\varepsilon=1/T$ along the modulation vector. Here, $T$ is the period of the magnetic structure in units of the lattice constant.

The results of neutron studies [44-50] of the magnetic texture of $La_{2-x}Sr_xCuO_4$ can be summed up in the form of a stripe phase diagram (Fig. 16a). It is seen that the incommensurate elastic-scattering peaks related to static modulation (shaded areas in Fig. 16a) are observed at $x \leq 0.07$. In the range $0.07 < x < 0.15$ incommensurate peaks are observed in inelastic neutron scattering, which are evidence of dynamic modulation of spin texture (open areas in Fig. 16a). At $x < 0.07$ there are one-dimensional "diagonal" stripes with a single modulation vector directed along orthorhombic axis b, while at $x > 0.05$ there is modulation in two directions parallel to the tetragonal axes ("parallel" stripes). To compare the spin structures for the cases of diagonal and parallel stripes, the diagonal stripes are examined in tetragonal units, too. Thus the spin modulation incommensurability parameter $d=\varepsilon$ for parallel and $d=\varepsilon/\sqrt{2}$ for diagonal stripes.

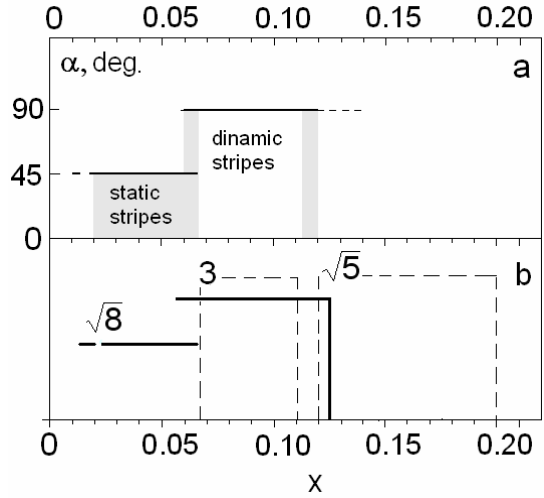

Fig. 16. Experimental magnetic phase diagram of $La_{2-x}Sr_xCuO_4$ [44-50] $\alpha=45^0$ and $90^0$ correspond to diagonal and vertical stripes, respectively. Hatched regions are the intervals where static stripes are observed. (b) Calculated stripe phase diagram of $La_{2-x}Sr_xCuO_4$. The dashed lines bound the regions of percolation along NUC chains with $l_{com}=3$ and $l_{com}=\sqrt{5}$ (dynamic stripes); the thick lines correspond to the regions of the existence of microdomains with doped holes ordered into a $\sqrt{8} \times \sqrt{8}$ lattice ($0.05 < x < 0.12$) and diagonal lines of doped holes ($x < 0.066$).

The theory, however, faces significant difficulties in describing the entire set of experimental results. The chief ones are:
1. the transition from diagonal to parallel stripes at $x \approx 0.05$;
2. the relation $d \approx x$ for $x<0.12$ and $d \approx const$ for $x>0.12$;
3. the change from static to dynamic stripes at $x \geq 0.07$ and emergence of static correlations within a narrow region of concentrations at $x \approx 0.12$ ("pinning of stripes");
4. the one-dimensionality of diagonal stripes and the two-dimensionality of parallel stripes;
5. the slant of parallel stripes.

In an attempt to explain the results of neutron experiment Gooding et al. [51,52] proposed the spin-glass model based on the supposition of chaotic distribution of localized doped holes. A localized doped hole is supposed to generates a long-range field of spin distortions of AFM background that can be described as the creation of a topological excitation, a skyrmion [53,54], with topological charge ±1 corresponding to twisting of AFM order parameter in the vicinity of a localized hole. Thus, doping destroys long range AFM order and leads to the formation of AFM-ordered microdomains whose angular points are specified by doped holes.

Combining some ideas from [51,52] together with our ideas concerning the mechanism of NUC formation and Sr ordering we provide below an alternative explanation of the observed spin and charge modulations.

**Parallel stripes.** Let us examine the case of complete ordering at $x_{com}=1/8$. We assume that each hole circulates over the oxygen sheet surrounding a copper ion and that because of the interaction between the hole current and the spins of the four nearest copper ions the latter are polarized. Figure 17 a shows a possible ordering of the projections of Cu spins on the $CuO_2$ plane for a completely ordered arrangement of localized holes at $x=1/8$. Here, the $CuO_2$ plane is broken up into separate AFM-ordered quadrangular microdomains whose corners are determined by the localized doped holes. The directions of Cu spin projections at lattice sites are indicated by arrows. The emerging pattern is characterized by the AFM-ordering of the microdomains proper and by the ordered alternation of skyrmions. Such a checkerboard pattern of AFM-ordered microdomains (Fig. 17a) results in an imitation of a "parallel stripe" structure. Actually, the "parallel stripes" are of the chains antiphase microdomains with magnetization directions along horizontal or vertical axes (Fig. 17a). The magnetic modulation period $T_m=8$ in this case is equal to the total size of two antiphase microdomains along the modulation vector and a rectangle with an area equal to $2 \times T_m$ must contain two sites. I.e.

$$2T_m x=2; \text{ and } d=1/T_m=x=1/8$$

Thus, the relation $d=x$ in the case of parallel stripes is caused exclusively by the fact that doped holes lie along straight lines equally spaced at $l_{com}=2$.

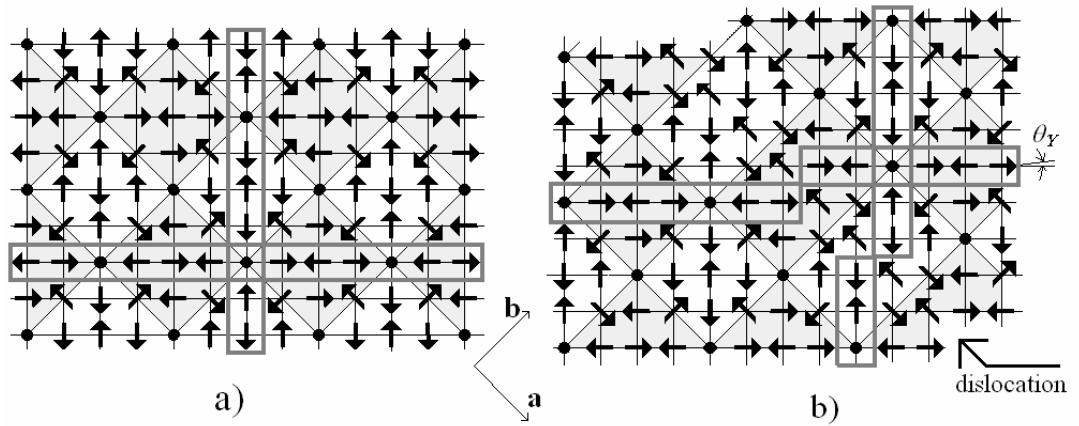

Fig. 17. (a) Projections of spin directions at $x = 1/8$ when doped holes are ordered into a $\sqrt{8} \times \sqrt{8}$ lattice. Microdomains that form horizontal stripes are hatched, thick lines denote stripe directions. (b) The same at $x < 1/8$. The plane is divided into domains separated by diagonal dislocations, which are nuclei of diagonal stripes. The shift of vertical stripes by one cell at each dislocation results in an effective tilt of parallel stripes by the $\theta_Y$ angle.

Now let us consider the magnetic textures formed at a slight deviation of $x$ below $x_{com}$=1/8. Kimura et al. [50] used an La$_{1.88}$Sr$_{0.12}$CuO$_4$ sample to observe the modulation of a spin texture with an incommensurability parameter $d$= 0.118. This corresponds to a mean texture period $\overline{T}_m \approx 8.5$ (in units of a), i.e., to the alternation of two periods, $T_{m1}$=8 and $T_{m2}$=9. Figure 17b shows the picture of an ordered distribution of doped holes we proposed for a mean concentration $x$=0.118, which was obtained by cutting the completely ordered lattice at $x$=1/8 along the orthorhombic $a$-axis and shifting one part in relation to the other by the vector q=(1,1). Such a dislocation conserves the coherence of ordering in domains on both sides of the dislocation (with a small phase shift). Such a structure (Fig. 17b) produces characteristic reflections in the diffraction pattern, and these reflections correspond to incommensurable modulation of both spin with an incommensurability parameter $d$. The condition of conservation of the mean concentration yields

$$T_d x_m = (T_d - 1) x_l$$

Here $T_d$ is the mean dislocation period in units of $a$. To maintain the mean concentration $x_m$=0.118 for a local concentration inside a domain $x_l$=0.125, the introduced diagonal dislocations must have a mean period $T_d$ =17 (=$T_{m1}$+$T_{m2}$). Such quasiperiodic dislocations, which lead to incommensurable modulation of the crystal structure and the spin texture, result in incommensurable reflections $d$ = 0.118 in full agreement with the experiment [50].

The special feature of the appearing picture of ordering is the shift of parallel stripes by one lattice constant (see Fig. 10b); that is, as though, they deflected from tetragonal axes by the $\theta_\gamma$ =1/17≈3.3° angle toward the orthorhombic axis $b$. These are those "slanted" parallel stripes with a slope angle of 3°, which were observed by Kimura et al. [50].

Let us now turn to the case of arbitrary values of $x$ for $x$<0.125. Here, the distribution pattern can be obtained from the completely ordered distribution at $x$=0.125 (Fig. 17a) by removing a certain number of lines of sites one after another. The texture imitating parallel stripes may occur down to $x$≈0.05 because small clusters (pieces of broken lines) with $l_{com}$=$\sqrt{8}$ still survive due to large relative fluctuations of concentration in small volume.

Let us suppose that the lattice contains such correlated remnant fragments of a parallel stripe texture genetically linked to $\sqrt{8} \times \sqrt{8}$ microdomains. The related neutron diffraction pattern exhibits characteristic reflections determined by the mean remnant texture period. In turn, the mean period $T$ of this texture, defined as the distance between the equivalent points of unidirectional magnetic domain, includes two occupied sites. I.e. a rectangle with an area equal to $T_m l_{com}/\sqrt{2}$ =2$T_m$ must contain two sites and $d=1/T_m=x$. This relation for parallel stripes takes place at 0.05<$x$<0.125, when the $\sqrt{8} \times \sqrt{8}$ lattice is occupied.

**Diagonal stripes.** As is seen from Figure 17b the inserted dislocations are actually the nucleus of diagonal stripes extended along $a$-axis. They appear as quasi-periodic structures at x<0,05 when $\sqrt{8} \times \sqrt{8}$ texture remainders disappear, and there only remain diagonal lines of impurity dipoles with a distance of $l_{com}$≥2$\sqrt{8}$ between the lines and a distance of $l_{com}$=$\sqrt{8}$ between the dipoles. Therefore diagonal stripes are always directed along the orthorhombic $a$-axes and, accordingly, the modulation vector, along the other orthorhombic $b$-axes. If all dipoles are ordered in diagonal ranks, the period of diagonal spin modulation $T_m$ (in tetragonal axes) should be equal $T_m$=1/$\sqrt{2}$ $x$ (or $d=\sqrt{2}$ $x$). Since the part of doped holes can remain in space between dipole ranks the period of observable spin structure will be more, than $1/\sqrt{2}\,x$, accordingly $d$ is less than $\sqrt{2}\,x$. The experimental value of $d$ varies [45] from $d \approx 0.7x$ up to $d \approx 1.4x$ over the range $0.01 < x < 0.05$.

**Dynamic stripes.** The last problem that we will discuss deals with dynamic stripes. Figure 16b shows the concentration ranges within which there can be antiferromagnetically correlated clusters of $\sqrt{8} \times \sqrt{8}$ microdomains ($0.05 < x < 0.125$) and diagonal chains of doped holes spaced $l = \sqrt{8}$ ($x < 0.07$). The dashed lines limit the regions of existence of percolation clusters with $l_{com} = 3$ and $l_{com} = \sqrt{5}$. Such conductive cluster bordering on AFM cluster destroys the static spin correlations because of the motion of charges that disrupt the magnetic order in the neighborhood along its path. Therefore spin correlations can be observed in the range $0.66 < x < 0.11$ only in inelastic neutron scattering as dynamic incommensurable magnetic fluctuations. What is remarkable (see Fig. 16b) is that in addition to the region $x < 0.07$ there is a narrow interval of concentrations $0.11 < x < 0.12$ where there is no percolation along NUC, and it is precisely in this interval that static incommensurable correlations are again observed (Fig. 16a). At nonhomogeneous Sr distribution the existence of small ordered domains with $x \leq 0.125$, bordering a percolative cluster with $l_{com} = \sqrt{8}$, is possible at $x > 0.125$. It results in the experimental observation of dynamic spin texture with $d \approx 0.125$.

**Conclusion.**

We have proposed the elementary model of high-$T_c$ superconductor based on the assumption that doped charges are localized, and their role is reduced to the corresponding local changes of volume Madelung energy and to the formation of NUC's on some pairs of Cu ions in $CuO_2$ plane. The latter is possible due to formation of a bound state (of the Heitler-London type) of two electrons on this pair of Cu ions and two holes on surrounding oxygen ions.

Electron pairing in such system, responsible for high-$T_c$ superconductivity, emerges because of strong renormalization of the effective electron-electron interaction when scattering with intermediate virtual bound states of NUC's is taken into account.

NUC's act as pair acceptors and the transitions of electron pairs from the oxygen band to the NUC's result in the generation of additional hole carriers. The calculation the Hall concentration for $YBa_2Cu_3O_{6+\delta}$ in dependence on temperature and doping has revealed the perfect agreement with experiment.

Two-particle hybridization of a pair level and states of the oxygen band leads to dramatically new properties of the system (nondegenerate distribution of hole carriers and the predominant contribution of electron-electron scattering to energy relaxation processes), which determine the unusual behavior of high-$T_c$ superconductors in the normal state including the temperature and frequency dependences of dc and optical conductivity.

The transitions o electrons on NUC's generate additional hole pairs, which are also localized in the vicinity of the NUC. Conduction occurs in such a system if these regions of hole localization form percolation clusters in $CuO_2$ plane (optimal doping state) and by means of quantum tunneling between such clusters (underdoping state). As well, conductivity can be provided if isolated NUC's are imbedded in metal matrix (overdoping state).

We have distinguished the structure fragments for $La_{2-x}Sr_xCuO_4$ where NUC's are formed and determined the concentration regions where these fragments form the percolation clusters of NUC's. We have shown that the specific ordering of dopant ordering takes place in $La_{2-x}Sr_xCuO_4$ that facilitates the formation of such percolation clusters.

We have found that the ordering of the dopant ions in La$_{2-x}$Sr$_x$CuO$_4$ in certain lattices leads to the formation of an incommensurate spin texture, which imitates stripe modulation, with an incommensurability parameter *d=x*.

We have discussed in detail the superconducting and magnetic phase diagrams of La$_{2-x}$Sr$_x$CuO$_4$ and have shown that the features of the phase diagrams only reflect the geometrical relations existing in a square lattice and the competition of different types of dopant ordering.

In the framework of the model the phase diagrams of YBa$_2$Cu$_3$O$_{6+\delta}$ is considered too. The interpretation of pseudogap and 60 K-phases in YBa$_2$Cu$_3$O$_{6+\delta}$ is offered. The pseudogap has superconducting nature and arises at temperature $T^* > T_{c\infty} > T_c$ in small clusters, uniting a number of NUC's, due to large fluctuations of NUC occupation. ($T_{c\infty}$ and $T_c$ are the temperatures of superconducting transition for infinite and finite clusters accordingly). The calculated $T^*(\delta)$ and $T_c(\delta)$ dependences are found to be in good accordance with experiment. The range between $T^*(\delta)$ and $T_c(\delta)$ corresponds to the range of fluctuations where small clusters fluctuate between superconducting and normal states owing to fluctuations of NUC occupation.

The close agreement between the calculated phase diagrams of La$_{2-x}$Sr$_x$CuO$_4$ and YBa$_2$Cu$_3$O$_{6+\delta}$ and the experimental results may serve as an important argument sustaining the proposed model of high-$T_c$ superconductor. As well, it is the decisive factor in favor of the mechanism of high-$T_c$ superconductivity in which superconducting pairing arises because of renormalization of electron-electron interaction due to scattering processes with intermediate virtual bound states of NUC's.

.